\documentclass[12pt,titlepage]{utarticle}
\usepackage{amsmath,amssymb,amsthm,dcpic,pictexwd,graphicx,color}
\usepackage{setspace}
\usepackage[mathscr]{eucal}
\usepackage[colorlinks=true, pdfstartview=FitV, linkcolor=blue, citecolor=blue, urlcolor=blue]{hyperref}

\numberwithin{equation}{section}

\def\defeq{\buildrel\rm def\over=}

\def\del{\partial}
\def\wdg{{\wedge}}                              % wedge product

\def\HH{{H\!H}}

\def\BC{\mathbb{C}}

\newcommand{\MR}[1]{{\mathbb{R}^{#1}}}            % Real numbers
\newcommand{\MC}[1]{{\mathbb{C}^{#1}}}            % Complex numbers
\newtheorem*{defn}{Definition}

\def\Spec{\mathrm{Spec}}

\def\Hom{\mathrm{Hom}}
\def\End{\mathrm{End}}

\def\Obj{\mathrm{Obj}}
\def\Arr{\mathrm{Arr}}
\def\Nat{\mathrm{Nat}}
\def\Ext{\mathrm{Ext}}

\def\dim{\mathrm{dim\ }}
\def\mcO{\mathcal{O}}
\def\mcA{\mathcal{A}}
\def\mcB{\mathcal{B}}
\def\mcC{\mathcal{C}}
\def\mcD{\mathcal{D}}
\def\mcE{\mathcal{E}}
\def\mcF{\mathcal{F}}

\def\mcM{\mathcal{M}}
\def\mcP{\mathcal{P}}
\def\id{\mathrm{id}}

\def\barQ{\overline{Q}}

\def\ie{\textit{i.e.,\ }}

\begin{document}
\preprint{UTTG--12--06\\
MIFP--06--24\\
\texttt{hep-th/0609225}\\}

\title{Deformations and D-branes}

\author{Aaron Bergman\address{George P. \& Cynthia W. Mitchell Institute for Fundamental Physics\\
             Texas A\&M University\\
             College Station, TX 77843-4242\\ {~}\\
             \email{abergman@physics.tamu.edu}}}
             
\Abstract{I discuss the relation of Hochschild cohomology to the physical states in the closed topological
string. This allows a notion of deformation intrinsic to the derived category. I use this to identify deformations of a quiver gauge theory associated to a D-branes at a singularity with generalized deformations of the geometry of the resolution of the singularity. An explicit map is given from noncommutative deformations (\ie B-fields) to terms in the superpotential.}

\maketitle
\newpage

\section{Introduction}\label{sec:intro}

The pioneering work of Douglas \cite{Douglas:2000gi} identified the open string information in the topological B-model with the derived category of coherent sheaves on the target space\footnote{Earlier works relating the derived category to string theory include \cite{Aspinwall:1998he,Kontsevich:1994ho,Sharpe:1999qz}.}. Each object in the derived category represents a boundary condition for the topological string, and the arrows between objects represent the physical states of an open string stretching between the two boundary conditions represented by the objects. This realization has had many exciting applications towards both physics and mathematics. The relevant one for this paper is the understanding of D-branes at a singularity.

Beginning with the works \cite{Cachazo:2001sg,Wijnholt:2002qz} and later elaborated in the works \cite{Aspinwall:2004vm,BridgeTStruct,Herzog:2003dj,Herzog:2004qw}, it was realized that the correspondence between D-branes at singularities and quiver gauge theories could, on the level of the topological string, be understood as an equivalence of categories between the derived category of coherent sheaves on a (crepant) resolution of the singularity and the derived category of representations of the quiver. While the physical import of this equivalence is not yet fully understood, it has many useful implications. It allows the easy identification of the fractional branes, an understanding of dibaryons \cite{Herzog:2003dj}, Seiberg duality \cite{Herzog:2004qw}, stability conditions \cite{Aspinwall:2004mb}, moduli spaces \cite{Bergman:2005kv,Bergman:2005mz,Wijnholt:2005mp} and more. In addition, recent work \cite{Hanany:2006nm} has connected this approach to the toric techniques pioneered in \cite{Feng:2000mi} and fully realized in \cite{Franco:2005rj,Franco:2005sm}.

The purpose of this paper is to take advantage of this correspondence to relate the deformations of the quiver gauge theory to the geometrical (and non-geometrical) deformations of the resolved\footnote{Note that the derived category is a birational invariant for Calabi-Yau 3-folds \cite{Bridgeland:2002fd}, so there is no ambiguity here.} singularity\footnote{Deformations of the singularity rather than just of the resolved geometry seem to be related to the addition of fractional branes to the theory. I will not attempt an analysis of that here. Examples have been worked out in \cite{Klebanov:2000hb,Berenstein:2005xa,Pinansky:2005ex}. Conifold-like transitions are also related to large-N dualities in the A-model \cite{Gopakumar:1998ki,Ooguri:2002gx}.}. The main tool we will use is the fact that closed string information is contained in the derived category. In fact, it has been well-known for some time among the mathematics community that the definition of the derived category due to Verdier is inadequate. It is best to replace it with either A$_\infty$-categories or differential graded (dg) categories. These turn out to be equivalent notions. A$_\infty$-categories may be familiar to the reader as the proper context for the Fukaya category of the A-model topological string. Homological mirror symmetry can then be understood as a (quasi-)equivalence between the Fukaya category and the derived category of coherent sheaves considered as A$_\infty$-categories.

The obvious question to ask is what is the extra information contained in the A$_\infty$ or dg structure. It is natural to conjecture that this is precisely the open string disc scattering amplitudes. That these amplitudes obey an A$_\infty$ structure was shown in \cite{Herbst:2004jp}. In addition, this can be seen from the point of view of holomorphic Chern-Simons theory in \cite{Aspinwall:2004bs}. Thus, these categories contain all the relevant information about the open string. However, any closed string can scatter onto a particular D-brane creating a string from that boundary condition to itself. We will see in the next section how we can identify the closed string states inside these enhanced derived categories. This is called the Hochschild cohomology ($\HH^\star$) of the category.\footnote{This observation seems to be well-known among some parts of the mathematical community (from where I learned it) and has appeared in some form in various physics papers (for example \cite{Kapustin:2004df}). It has been suggested in \cite{Moore:toap} that cyclic cohomology might be a better answer, but I will not address that issue.} The Hochschild-Kostant-Rosenberg (HKR) theorem (as generalized in \cite{Kontsevich:2003dq,Swan:1996hc,Yekutieli:2002ab}) states that, for a quasiprojective algebraic variety, $X$,
\begin{equation}
\label{HKR}
\HH^i(\mcD(X)) \cong \bigoplus_{j+k=i} H^j(X,\wedge^k TX)\ .
\end{equation}
which the reader will recognize as the space of closed string states in the B-model.

Once we have identified the closed string states, we can identify the infinitesimal deformations of our theory. We can consider any dimension two operator as a candidate to exponentiate and add to the action. The descent process relates the group $\HH^2$ to the space of infinitesimal deformations (ignoring issues with the $U(1)$ charges \cite{Kapustin:2004gv}). There are obstructions to extending these deformations to higher order which are elements in $\HH^3$. Using the decomposition \eqref{HKR}, we identify $H^1(X,TX)$ as the complex structure deformations and $H^0(X,\wedge^2 TX)$ as the noncommutative deformations given by a B-field. The final group $H^2(X,\mcO)$ is a sort of gerbey deformation that we will not discuss. It may perhaps be better to think of these deformations as arising in the context of generalized complex geometry as is demonstrated in \cite{Gualtieri:2004gc}.

The equivalence of categories for D-branes at a singularity in the B-model extends to an equivalence of dg-categories \cite{Toen:2004to}. Thus, given the intrinsic definition of Hochschild cohomology to the derived category (which will, henceforth, always refer to the suitably enhanced category), we can describe the geometrical deformations given by the HKR theorem above in terms of the quiver gauge theory. Unsurprisingly given its name, this Hochschild cohomology for the derived category of representations of the quiver algebra is precisely the usual Hochschild cohomology of the algebra. I will discuss the well-known fact that $\HH^2$ for the algebra describes infinitesimal deformations of the algebra. I will show how superpotential deformations of the quiver gauge theory appear as elements in the Hochschild cohomology and describe, in principle, how to match them with deformations of the geometry. In addition, I will, in many cases, give an explicit map between the noncommutative deformations of the geometry and terms in the superpotential.

The specific result I obtain is as follows. Consider a del Pezzo surface, $X$, such that the group $H^1(TX \otimes \omega_X^{-p})$ vanishes\footnote{This holds for all del Pezzos without complex structure deformations, for example.} for all $p > 0$. From this, we construct a quiver gauge theory corresponding to placing a D-brane at the tip of the canonical cone $\omega_X$ with the zero section collapsed. The choice of an exceptional collection gives an ordering of the nodes of the quiver. The space of B-fields contains $\bigoplus_{p\ge0}H^0(\omega_N^{-p})$, and we have a map from this group into the space of loops at a given node. For the node $n$ (corresponding to the rightmost element in the exceptional collection as I will describe), we will see that this loop is precisely the deformation of the superpotential. It would be interesting to perform this matching for situations where there are complex structure deformations.

I will use the language of derived categories and derived functors extensively in this paper. For a nice introduction for physicists, see \cite{Aspinwall:2004jr}. For the general theory, the textbooks \cite{GelMan,Weibel} are invaluable. The relation between Hochschild cohomology, deformations and string theory has been discussed from a different perspective in \cite{Hofman:2000ce,Hofman:2001zt}. Noncommutative deformations of the moduli space of quiver gauge theories in the context of D-branes at a singularity have been examined in \cite{Wijnholt:2005mp}.

This paper is organized as follows. In section two, I will discuss, following \cite{Moore:toap}, closed string states in general 2D topological field theories and show how they relate to Hochschild cohomology. In section three, I will discuss how Hochschild cohomology maps to deformations of the geometry as in the HKR theorem above, and how it relates to deformations of algebras. In section four, I will briefly cover the equivalence of categories that is at the heart of understanding the relation between quiver gauge theories and D-branes at singularities. In section five, I will review relevant information from the homological algebra of quivers and show how to construct the quiver gauge theory. In section six, I will show how superpotential deformations are elements in $\HH^2$ of the quiver algebra. Finally, in section seven, I will give the explicit map between noncommutative deformations  and certain superpotential deformations.

\section{Hochschild Cohomology and 2D TFT}\label{sec:2dtft}

\subsection{The Moore-Segal category}

As originally formulated by Atiyah \cite{Atiyah:1988at} (following work of Segal \cite{Segal:1988ft,Segal:2004cf,Segal:1981se}), a topological quantum field theory of dimension $d$ is given by a (symmetric monoidal) functor from the bordism category of  $d$-dimensional manifolds to the category of vector spaces. For a nice introduction, see \cite{Dijkgraaf:1997ip}. This can be generalized to an open-closed TQFT by allowing labelled boundaries in addition to the `in' and `out' boundaries on the manifolds in the bordism category \cite{Moore:toap,Moore:lect,Lauda:2005wn}.

\begin{figure}[t]
\begin{center}
\includegraphics{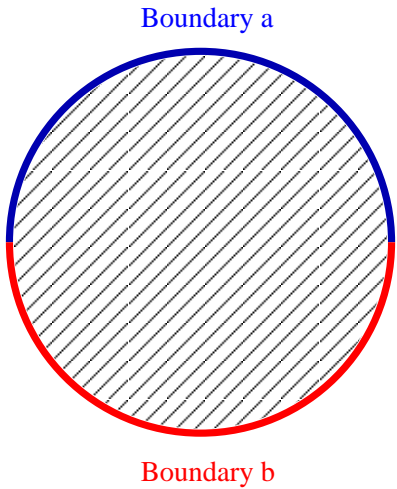}
\end{center}
\caption{An open string between two D-branes.}
\label{2bdrystring}
\end{figure}

\begin{figure}[t]
\begin{center}
\includegraphics{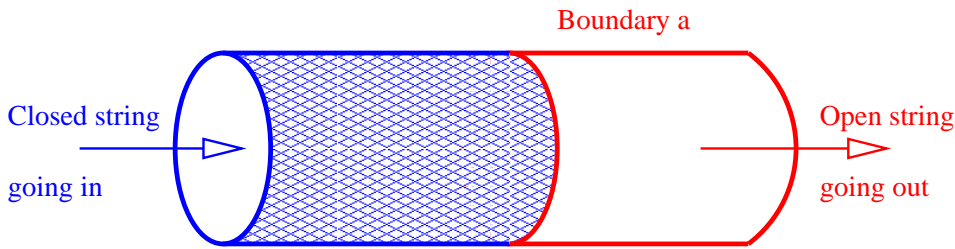}
\end{center}
\caption{A closed string splitting into an open string.}
\label{closedtoopen}
\end{figure}

\begin{figure}[t]
\begin{center}
\includegraphics{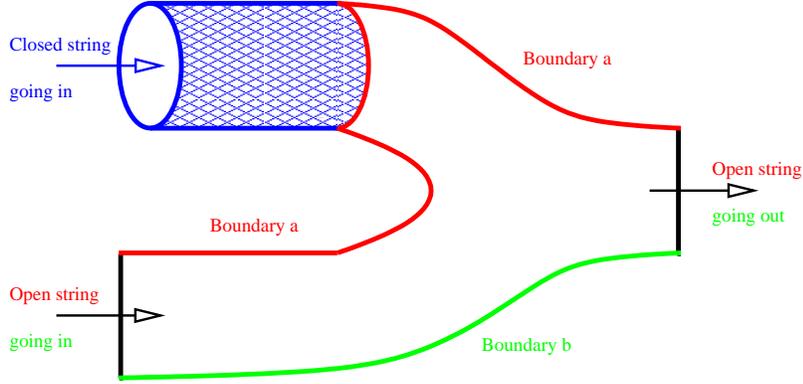}
\end{center}
\caption{The composition of $i_a$ with the open string product.}
\label{prodclosed}
\end{figure}

Given an open-closed TFT, we can form a category, $\mcC$, with the boundary labels as objects and the arrows being the vector spaces corresponding to the physical states of the diagram in figure \ref{2bdrystring}. The closed strings form a commutative algebra which we will denote $A$. Given a particular boundary condition, $a \in \Obj(\mcC)$, we have the noncommutative algebra $\End_\mcC(a) = \Hom_\mcC(a,a)$. The diagram in figure \ref{closedtoopen} maps closed string states to open string states $i_a : A \rightarrow \End(a)$. The composition of  \ref{closedtoopen} with the open string product gives figure \ref{prodclosed}. This can be deformed to the same figure, but with the incoming open string on top, giving
\begin{equation}
\label{ctooprel}
x i_a(\alpha) = i_b(\alpha) x
\end{equation}
where $x$ is a string between the $b$ and $a$ boundary conditions and $\alpha$ is a closed string state. In particular, when $a=b$, this implies that the image of $i_a$ lies in the center of $\End(a)$.

This condition is natural in category theory. To understand this, we will need the notion of a \textit{natural transformation}:

\begin{defn}
Given two categories $\mcA$ and $\mcB$, and two functors between them, $F$ and $G$, a \textbf{natural transformation} between $F$ and $G$ is a map $\eta : \Obj(\mcA) \rightarrow \Arr(\mcB)$ such that, for $x \in \Obj(\mcA)$, we have $\eta(x) \in \Hom_\mcB(F(x),G(x))$, and for $x,y \in \Obj(\mcA)$ and $a \in \Hom_\mcA(x,y)$, the following diagram commutes:
\end{defn}
\begin{equation}
\label{nattrans}
\begin{split}
	\begindc{\commdiag}[3]
		\obj(10,10)[Gx]{$G(x)$}
		\obj(30,10)[Gy]{$G(y)$}
		\obj(10,30)[Fx]{$F(x)$}
		\obj(30,30)[Fy]{$F(y)$}
		\mor{Fx}{Fy}{$F(a)$}
		\mor{Gx}{Gy}{$G(a)$}
		\mor{Fx}{Gx}{$\eta(x)$}[\atright,\solidarrow]
		\mor{Fy}{Gy}{$\eta(y)$}
	\enddc
\end{split}\ .
\end{equation}

A natural transformation is a sort of map between functors. Just as it is better to speak of isomorphisms between objects in a category rather than equality between objects, it is better to talk of natural transformations between functors rather than equality of functors. For example, two categories, $\mcA$ and $\mcB$, are considered \textit{equivalent} if there are functors $F : \mcA \to \mcB$ and $G : \mcB \to \mcA$ such that there are natural transformations $FG \rightarrow \id_\mcB$ and $\id_\mcA \to GF$ whose corresponding maps of objects are isomorphisms (these are called natural isomorphisms).

Recall now our category of boundary conditions, $\mcC$. We have maps $i_a : A \to \End(a)$ for all objects $a \in \mcC$. It is not hard to see that this is exactly a natural transformation from the identity functor to itself\footnote{I learned of this from a lecture by Dan Freed on the work of Moore and Segal \cite{Moore:toap}.}. The diagram \eqref{nattrans} exactly encodes the relation \eqref{ctooprel}. Thus, we have an embedding of the closed string algebra, $A \hookrightarrow \Nat(\id_\mcC,\id_\mcC)$. This latter space is a toy version of Hochschild cohomology.

\subsection{Cohomological TQFTs}

The topological string is an example of a cohomological TQFT \cite{Witten:1990bs} in that the underlying QFT is not topological, but once we pass to the cohomology of a BRST-like operator, the physical states and amplitudes are topological. For the topological string, the underlying QFT is a conformal theory. This has led mathematicians to term these theories topological conformal field theories (TCFTs).

These were axiomatized by Segal \cite{Segal:1999tf} and Getzler \cite{Getzler:1994yd} as follows. We begin with Segal's category of Riemann surfaces which we denote $\mcM$. This is similar to the bordism category used above, except that instead of manifolds with boundaries, we use Riemann surfaces with boundaries as the morphisms. A conformal field theory, as per Segal \cite{Segal:2004cf}, is a (symmetric monoidal) functor from this category to the category of vector spaces. Now let $\mcC_*$ be the functor that assigns to a topological space the complex of singular chains. By applying this to Segal's category of Riemann surfaces, we obtain a differential graded (dg) category, $\mcC_*(\mcM)$. As with the Segal category, the objects are still the natural numbers, but the morphisms are given by chains on the moduli space of Riemann surfaces with the specified boundaries. Thus, each Hom-space is a graded vector space along with a degree one operator that squares to zero. A TCFT is, then, a (symmetric monoidal) dg-functor from $\mcC_*(\mcM)$ to dg-Vect, the category of differential graded vector spaces, \ie chain complexes. (A dg-functor is a functor that respects the dg structure on the Homs.) With a little thought, one can see how this codifies the structures found in the topological string.

We can now see how a dg-category (or more generally an A$_\infty$-category) can arise from a TCFT. As above, choose a set of boundary conditions and, for a pair of boundary conditions, let the Hom-space between them be given by the complex (\ie dg-vector space) given by the string states between them with the action of a BRST operator. A new feature here is the shift functor, denoted $[n]$ that takes a boundary condition to itself with a shifted grading. The physical origin of this functor is given by Douglas in \cite{Douglas:2000gi}. With this in hand, we can define Hochschild cohomology for a dg-category, $\mcB$:
\begin{equation}
\label{HHdef}
\HH^i(\mcB) = \Nat(\id_\mcB, [i])\ .
\end{equation}

Costello \cite{Costello:2004ei} made this precise by extending the definition of TCFTs to open-closed theories. He proves that, given a set of boundary conditions, the category of open-closed TCFTs with those boundary conditions is roughly equivalent to the category of Calabi-Yau A$_\infty$-categories. (Please see the reference for a precise statement of the results.) He then constructs a map from the Hochschild homology (which I will not define) of this category to the closed string states. In addition to \eqref{HKR}, the HKR theorem also states that Hochschild homology is isomorphic to $H^*(\Omega^*)$ which is isomorphic to Hochschild cohomology as given in \eqref{HKR} for a Calabi-Yau variety. This can be understood as follows \cite{Kapustin:2004gv}. In the untwisted $(2,2)$ sigma model, the space of ground states is given by the RR-forms which we see is the Hochschild homology. In addition, the chiral ring is precisely the Hochschild cohomology. In the case of the topological sigma-model, the state-operator correspondence gives the isomorphism between them, and in the (2,2) sigma-model it can also be interpreted as spectral flow. 

\subsection{The dg-derived category}

In this paper, we will deal solely with the topological B-model. While A$_\infty$-categories are, in a useful sense, equivalent to dg-categories, a dg-category equivalent to a given A$_\infty$ category can be quite complicated. Thankfully, for the B-model (unlike the A-model) such a nice description exists\footnote{I am not aware of a good reference for this material. I would like to thank David Ben-Zvi for explaining it to me.}.  In particular, for target space $X$ we can form the category $\mcD(X) \defeq \mcD(Coh(X))$ where the objects are complexes of vector bundles\footnote{It is not necessary to include all coherent sheaves as on a smooth variety any coherent sheaf can be resolved to a complex of vector bundles, \ie locally free sheaves.} and the Hom-spaces are given as follows. Let $\mcE^\star$ and $\mcF^\star$ be complexes of vector bundles. Then, $\Hom^0(\mcE^\star,\mcF^\star)$ is given by maps $f^i : \mcE^i \to \mcF^i$ as in the following diagram:
\begin{equation}
\label{homspace}
\begin{split}
\begindc{\commdiag}[3]
	\obj(10,10)[lb]{}
	\obj(30,10)[F3]{$\mcF^3$}
	\obj(50,10)[F2]{$\mcF^2$}
	\obj(70,10)[F1]{$\mcF^1$}
	\obj(90,10)[rb]{}
	\obj(10,30)[lt]{}
	\obj(30,30)[E3]{$\mcE^3$}
	\obj(50,30)[E2]{$\mcE^2$}
	\obj(70,30)[E1]{$\mcE^1$}
	\obj(90,30)[rt]{}
	\mor{lb}{F3}{}
	\mor{F3}{F2}{}
	\mor{F2}{F1}{}
	\mor{F1}{rb}{}
	\mor{lt}{E3}{}
	\mor{E3}{E2}{}
	\mor{E2}{E1}{}
	\mor{E1}{rt}{}
	\mor{E1}{F1}{$f^1$}
	\mor{E2}{F2}{$f^2$}
	\mor{E3}{F3}{$f^3$}
\enddc
\end{split}
\end{equation}
Note that we do \textit{not} require that the above diagram commutes. We can generalize this to the graded vector space $\Hom^i_{\mcD(X)}(\mcE^\star,\mcF^\star) \defeq \Hom^0_{\mcD(X)}(\mcE^\star,\mcF^\star[i])$. Abusing the notation to let $\delta$ denote the differential on both $\mcE^\star$ and $\mcF^\star$, we can define a differential on this Hom-space by $d(f) = [\delta,f]_\pm$ where that is a graded commutator. It is straightforward to see that this squares to zero (taking into account the grading), and we have thus defined a dg-category. We then invert the quasi-isomorphisms to obtain our dg-enhancement of the usual derived category.\footnote{In general, we may have to consider twisted complexes or perhaps pass to the saturation of the category (which includes the twisted complexes). The physical import of twisted complexes is explained in \cite{Diaconescu:2001ze}.} Note that the closed maps are precisely the chain maps between $\mcE^\star$ and $\mcF^\star$, and the exact maps are homotopic to zero. To save space, we will denote
\begin{equation}
\Hom_{\mcD(X)}(\mcE^\star,\mcF^\star) \defeq H^0(\Hom^\star_{\mcD(X)}(\mcE^\star,\mcF^\star))\ .
\end{equation}
Recall that in the usual construction of the derived category, we begin with chain maps and pass to the homotopy category where the maps are homotopy classes of chain maps. We see now that this is exactly the cohomology of our category, so $\Hom_{\mcD(X)}(\mcE^\star,\mcF^\star)$ is the space of maps in the usual derived category. By avoiding the passage to cohomology, we have retained further information. As mentioned in the introduction this extra information should precisely be the open string scattering amplitudes. This dg generalization of the derived category can be defined similarly for any abelian category such as the category of modules over an algebra.

\section{Hochschild cohomology and deformations}\label{sec:hochdef}

\subsection{Algebra deformations}

In this section, we will relate Hochschild cohomology as defined in \eqref{HHdef} to the traditional definition of Hochschild cohomology for algebras. Let $A$ and $B$ be algebras. We will denote $\mcD(A) \defeq \mcD(A \mathrm{- mod})$ and similarly for $B$. The theory of derived Morita equivalence due to Rickard \cite{Rickard:1988mt} and extended to dg-categories by To\"en \cite{Toen:2004to} tells us that any functor between the two categories, $F : \mcD(A) \to \mcD(B)$, is equivalent to the existence of an $A - B$ bimodule $M$ with $F(X) = X \otimes^\mathbf{L}_A M$. $A$ considered as a bimodule over itself corresponds to the identity functor. This allows us to write \eqref{HHdef} as
\begin{equation}
\label{HHalgdef}
\HH^i(A) \defeq \HH^i(\mcD(A)) = \Hom_{\mcD(A - A)}(A,A[i]) = \Ext^i_{A - A}(A,A)\ .
\end{equation}
where $A-A$ is the abelian category of $A$ bimodules.

In order to compute the Ext group, we need a resolution of $A$ as a bimodule. There is a canonical projective resolution called the bar resolution. It is defined as follows (see, for example, \cite{MR2002c:55027}). We will consider $A$ as an algebra over a field, $k$. Let $A^{\otimes n} = A \otimes_k A \otimes_k \dots \otimes_k A \otimes_k A$ for a total of $n$ copies of $A$. These are $A$ bimodules with the bimodule structure given by the multiplication by $A$ on the left and right copies of $A$. It is not hard to see that these are projective bimodules. We will denote an element of $A^{\otimes n}$ as $[a_1|a_2|\dots|a_n]$. The vertical bars are the reason this is called the bar resolution.

We now have the following exact sequence of projective bimodules:
\begin{equation}
\label{barres}
\dots \longrightarrow A^{\otimes 4} \longrightarrow A^{\otimes 3} \longrightarrow A \otimes_k A \longrightarrow A \longrightarrow 0\ .
\end{equation}
The map from $A^{\otimes n} \to A^{\otimes (n-1)}$ is given by
\begin{equation}
\label{barmap}
[a_1|a_2|\dots|a_n] \mapsto \sum_i (-1)^i [a_1 | \dots | a_{i} a_{i+1} | \dots | a_n]\ .
\end{equation}
The sequence without the final $A$ is then a projective resolution of $A$.

To compute $\Ext^i_{A - A}(A,A)$ we apply the functor $\Hom_{A-A}(-, A)$ to the resolution given by \eqref{barres}. It is straightforward to verify that, for a $k$-vector space $T$,
\begin{equation}
\label{freemap}
\Hom_{A-A}(A \otimes_k T \otimes_k A, A) \cong \Hom_k(T, A)
\end{equation}
where the latter Hom is as vector spaces. Thus we have that $\Hom_{A-A}(A^{\otimes n},A)$ is isomorphic to $\Hom_k(A^{\otimes(n-2)}, A)$, and $\Ext^i_{A-A}(A,A)$ is given by the $i$-th cohomology of the following sequence:
\begin{equation}
\label{oldhoch}
A \longrightarrow \Hom(A,A) \longrightarrow \Hom(A\otimes_k A, A) \longrightarrow \Hom(A^{\otimes 3},A) \longrightarrow \dots\ .
\end{equation}
The differential is given by:
\begin{equation}
\label{oldhochmap}
\begin{split}
(\delta f)(a_1,\dots,a_{n+1}) = &a_1 f(a_2,\dots,a_{n+1}) + \sum_i (-1)^i f(a_1,\dots,a_i a_{i+1},\dots,a_{n+1}) \\ 
&+ (-1)^{n+1}f(a_1,\dots,a_n)a_{n+1}\ .
\end{split}
\end{equation}
This is the traditional definition of Hochschild cohomology for algebras (see, for example, \cite{Ginzburg:2005nc}).

The first few Hochschild cohomology groups have well-known interpretations. The zeroth group $\HH^0(A) = Z(A)$, the center of $A$. It is not too hard to see that $\HH^1(A)$ is precisely the outer derivations of $A$. Finally, $\HH^2(A)$ is the space of first order formal deformations of A as we will now see. For a nice overview of this material, see \cite{Etingof:2005nc}.

As we do with the Moyal-Weyl star product in noncommutative gauge theories, we will define a formal deformation of an algebra by deforming the product on that algebra. Denote the product on a $k$-algebra $A$ by $m : A \otimes_k A \to A$. Then, a formal deformation of the product is given by a formal power series, \ie a map $\mu : A \otimes_k A \to A[[\hbar]]$. Note that we are just using $\hbar$ as a formal parameter here; it has nothing to do with the physical $\hbar$. This gives a set of functions $\mu_i : A \otimes_k A \to A$ for $i = 1 \dots \infty$ as follows:
\begin{equation}
\label{defprod}
\mu(a,b) = m(a,b) + \sum_{i=1}^\infty \hbar^i \mu_i(a,b)\ .
\end{equation}

Truncating to first order in $\hbar$, associativity implies that $\delta\mu_1 = 0$ where $\delta$ is as in \eqref{oldhochmap}. If two $\mu_1$ differ by a coboundary, \ie a $\delta f$ for any $f : A \to A$, then the deformed products are equivalent to first order. Thus, first order formal deformations are in exact correspondence to $\HH^2(A)$. It turns out that the higher conditions for associativity all reduce to elements in $\HH^3(A)$. When these vanish, the deformation can be extended to higher orders. In particular, if $\HH^3(A) = 0$, then all first deformations can be extended to formal deformations. Later, we will apply this technology to the algebras obtained from quiver gauge theories and see how superpotential deformations give rise to elements in $\HH^2$.

\subsection{Geometrical deformations}\label{ssec:geo}

In this section, we will discuss the Hochschild cohomology of the derived category of coherent sheaves on a variety (or, more generally a quasiprojective scheme). A nice reference for some of this material (and a general review of derived categories) is \cite{Caldararu:2005dc}.  We will begin with the fact that all nice enough functors between the derived category of quasicoherent sheaves on two varieties, $X$ and $Y$, are given by Fourier-Mukai transforms \cite{Toen:2004to}. Recall that a Fourier-Mukai transform is given by an object $\mcM \in \Obj(\mcD(X \times Y))$. Then, the corresponding functor $F : \mcD(Y) \to \mcD(X)$ is given by
\begin{equation}
\label{FMT}
\mcA \in \Obj(\mcD(X)) \mapsto F(\mcA) = \mathbf{R}\pi_{1*}(\mcM \otimes^\mathbf{L} \pi_2^*(\mcA))
\end{equation}
where $\pi_1$ and $\pi_2$ are the projections from $X \times Y$ to $X$ and $Y$ respectively. $\mcM$ is often called the kernel of the transform.

Let $\Delta : X \hookrightarrow X \times X$ be the diagonal embedding for X. Then, the identity functor from $\mcD(X)$ to itself is given by the kernel $\mcM = \Delta_*\mcO_X = \mcO_\Delta$ which is a coherent sheaf. This allows us to rewrite the functorial definition of Hochschild cohomology \eqref{HHdef} as
\begin{equation}
\label{HHspacedef}
\begin{split}
\HH^i(X) & \defeq \HH^i(\mcD(X)) = \Hom_{\mcD(X \times X)}(\mcO_\Delta,\mcO_\Delta[i]) = \Ext^i_{X\times X}(\mcO_\Delta,\mcO_\Delta) \\
&=\Hom_{\mcD(X)}(\mathbf{L}\Delta^*\mcO_\Delta,\mcO_X[i]) = \Ext_X^i(\mathbf{L}\Delta^*\mcO_\Delta,\mcO_X)\ .
\end{split}
\end{equation}

The original HKR theorem \cite{Hochschild:1962df} states that, for a commutative algebra, $A$, $\HH^i(A) \cong  \wedge^i\mathrm{Der}(A)$ where Der$(A)$ is the space of derivations of A. This is equivalent to the statement that $\HH^i(\Spec(A)) \cong H^0(\Spec(A),\wedge^i T\Spec(A))$. As any variety can be covered by affine patches, we can think of the HKR theorem as a globalization of this result. In particular, for a quasiprojective variety, $X$, we have from Kontsevich \cite{Kontsevich:2003dq}, Swan \cite{Swan:1996hc} and Yekutieli \cite{Yekutieli:2002ab}:
\begin{equation}
\label{HKR2}
\HH^i(\mcD(X)) \cong \bigoplus_{j+k=i} H^j(X,\wedge^k TX)\ .
\end{equation}

This isomorphism is accomplished by something called the Atiyah-Chern character \cite{Buchweitz:2006gd,Caldarary:2005hk,Markarian:2001pb,Ramadoss:2006rr,Ramadoss:2005bc},
\begin{equation}
\label{ACgroup}
AC \in \Hom_{\mcD(X\times X)}\left(\mcO_\Delta, \bigoplus_{i=0}^{\dim X} \Omega^i X [i]\right)\ .
\end{equation}
This gives rise to a quasiisomorphism:
\begin{equation}
\label{HKRiso}
I : p^*\left(\bigoplus_{i=0}^{\dim X} \wedge^i TX [-i]\right) \otimes \Delta_*\mcO_X {\buildrel\sim\over\longrightarrow} \mathscr{E}xt_{X \times X}(\mcO_\Delta,\mcO_\Delta)
\end{equation}
where $p$ is either projection from $X \times X \to X$. The isomorphism \eqref{HKR2} immediately follows.

We will not use this form of the isomorphism here, however. Instead we will compute $\mathscr{E}xt_{X \times X}(\mcO_\Delta,\mcO_\Delta)$ and look at its cohomology with other techniques. It should be noted that this isomorphism does not respect the product structure on Hochschild cohomology. This can be remedied by composing $I$ with a version of the Todd genus \cite{Markarian:2001pb,Ramadoss:2006rr,Ramadoss:2005bc}.

We immediately see from \eqref{HKR2} that $\HH^2(X)$ contains $H^1(X,TX)$, the space of infinitesimal complex structure deformations. The other two groups can be considered as generalized deformations. $H^0(X,\wedge^2 TX)$ is a global bivector, for example, giving rise to a noncommutative deformation. Recall that, in the Seiberg-Witten limit, we take the matrix  inverse of the B-field to obtain the noncommutative deformation $\theta$ \cite{Seiberg:1999vs}. That theta is precisely this bivector. (Mathematically, this is equivalent to inverting a symplectic form to obtain a Poisson structure.) It is a theorem of Kontsevich \cite{Kontsevich:2003dq} that deformations given by a Poisson structure are unobstructed. The final group $H^2(X,\mcO)$ is a gerbey deformation and still physically somewhat mysterious. This group vanishes in all the situations that will arise in this paper and also on all compact Calabi-Yaus with holonomy $SU(3)$ (as opposed to some subgroup). While it is possible, in principle, to explicitly match all these deformations with superpotential deformations, in this paper we will focus on the noncommutative deformation.

\section{D-branes and derived Morita equivalence}\label{sec:dbranes}

\subsection{The equivalence}

In this section, I will discuss the equivalence of categories that is at the heart of understanding D-branes at a singularity. As this material is covered extensively elsewhere, I will omit some information and adopt a perspective slightly different from that in other physics papers. This material is mostly drawn from \cite{BridgeTStruct}.

The basic geometry we will deal with is type IIB string theory compactified on $\MR{1,3} \times M$ where $M$ is a Calabi-Yau 3-fold. We will assume that $M$ has a singularity and that there is a D3-brane filling the $\MR{1,3}$ and located at the singularity. We want to determine the gauge theory that lives on the D-brane. The idea is that this brane is marginally stable to decay into a set of `fractional' branes. Associated to each fractional brane is a $U(N)$ gauge group, and the string states between the fractional branes give bifundamental matter. This is the data of a quiver gauge theory.

Much of this problem can be analyzed in the context of the topological B-model. We will be able to identify possible sets of fractional branes and associate quiver gauge theories to all of them. The `correct' set of fractional branes depends on the question of marginal stability and, as such, depends on the K\"ahler information.

The main tool we will use is the description of the derived category of coherent sheaves as a derived category of representations of an algebra. In particular, we will find an object that generates the derived category, by which we mean that every object in the derived category is quasiisomorphic to a complex that only involves direct sums of our generating object. If we have such an object, $T$, that satisfies
\begin{equation}
\label{homcond}
\Hom(T,T[n]) = 0 \text{\quad for } n \neq 0\ .
\end{equation}
then it follows from the theory of derived Morita equivalence \cite{Rickard:1988mt} that we have an equivalence of categories $\mcD(M) \cong \mcD(\End(T)^\text{op} \mathrm{- mod})$. Let $A \defeq \End(T)^\text{op}$. Since we are looking for a set of fractional branes into which the original brane can decay, we will look for objects, $T$, which are decomposable. In other words, we desire that $T \cong \bigoplus E_i$. 

Since we are only looking to describe the D-brane at the singularity, we will `zoom in' and consider conical Calabi-Yaus. In particular, let $N$ be a smooth Fano surface. More generally, one should consider orbifolds and possibly even noncommutative spaces. Let $M$ be the total space of the canonical line bundle on $N$. Thus, $M$ has trivial canonical class. We will assume that a Calabi-Yau metric exists on $M$ (where we consider $M$ as a complex variety, not just a topological manifold). This is true for all smooth Fano surfaces (the del Pezzo surfaces) and for a large class of orbifolds \cite{Boyer:2005sg}. One can also describe other line bundles that give rise to Calabi-Yau metrics off the zero section \cite{Bergman:2005ba}, but we will not address this here.

We will proceed (following \cite{BridgeTStruct}) by first constructing an equivalence of categories for $\mcD(N)$ and then lifting to an equivalence of categories for $\mcD(M)$. Bondal \cite{BondalQuiv} tells us how to construct this equivalence to the derived category of representations of a quiver algebra by means of a strong exceptional collection. This is a collection of coherent sheaves $E_i$ such that
\begin{eqnarray}
\label{exccond1}
\Ext_N^i(E_a,E_b) &=& 0 \text{\quad for } i \neq 0\ , \\
\label{excond2}
\Hom_N(E_a,E_b) &= &0 \text{\quad for } a > b\ , \\
\label{excond3}
\Hom_N(E_a,E_a) &=& \BC \ .
\end{eqnarray}
The first condition ensures that \eqref{homcond} holds. As above, let $T = \bigoplus E_i$ and $A = \End(T)^\text{op}$. The decomposition of $T$ tells us that the identity element of $A$ decomposes into a sum of idempotents. This defines for us a quiver algebra (with relations) where each node corresponds to an idempotent. The second two conditions (\ref{excond2},\ref{excond3}) tell us that this quiver algebra has no loops. (For more details, please see \cite{BondalQuiv}.) Since $T$ maps to the free module over $A$, we have that the $E_i$ map to the projective representations of $A$. Using this identification, we will see later that $\Hom(E_i,E_j)$ is isomorphic to the space of paths from node $j$ to node $i$.

Now, let $\pi$ denote the projection from $M \to N$. Then, we can consider the object $\pi^*T$. It generates $\mcD(M)$ if we have that $\Hom_{\mcD(M)}(\pi^*T,\mcE) = 0 \Longrightarrow \mcE = 0$. This can be seen by applying the adjunction:
\begin{equation}
\label{genadj}
\Hom_{\mcD(M)}(\pi^*T,\mcE) \cong \Hom_{\mcD(N)}(T,\pi_*\mcE) = 0\ .
\end{equation}
Since $T$ generates $\mcD(N)$, this implies that $\pi_*\mcE = 0$. As $\pi_*$ is an exact functor with no kernel, we have $\mcE = 0$. Next, we need to ensure that $\pi^*T$ satisfies \eqref{homcond}. This gives\begin{equation}
\Hom_M(\pi^*T,\pi^*T[i]) = \Hom_N(T,\pi_*\pi^*T[i]) = \bigoplus_{n \geq 0} \Hom_N(T, T \otimes \omega_N^{-n}[i]) = 0 \quad \text{for } i \neq 0\ .
\end{equation}
or
\begin{equation}
\label{simphel}
\Hom_N(E_a,E_b \otimes \omega_M^p[i]) = 0 \quad \text{for } i\neq 0, p \leq 0\ .
\end{equation}
This condition is very similar to the definition of a geometric helix in \cite{BondPol}. We will say that the $E_i$ generate a \textit{simple helix}, but this terminology is not standard. This gives us the desired equivalence of categories and is Proposition 4.1 of \cite{BridgeTStruct}. In the next section, we will see how to describe $B \defeq \End(\pi^*T)^\text{op}$ in terms of a quiver. This will be the quiver of a quiver gauge theory that can describe the D-brane at the singularity obtained by collapsing the zero-section of $M$ to a point.

Similar to the above, the $\pi^*E_i$ map to projective representations of $B$, and we will have that paths from node $j$ to node $i$ are given by
\begin{equation}
\label{loopshom}
\Hom_M(\pi^*E_i,\pi^*E_j) \cong \bigoplus_{p \ge 0} 
	\Hom_N\left(E_i, E_j \otimes \omega^{-p}_N\right)\ .
\end{equation}
These are infinite dimensional vector spaces, but each piece graded by $p$ is finite dimensional. In the quiver, $p$ will translate to the number of times the path circles the quiver. This is well-defined given the ordering of the nodes.

\subsection{Discussion}

Recall that a Calabi-Yau category of dimension $d$ is one in which the Serre functor (see \cite{Bondal:1989rf}) is equivalent to the functor $[d]$. For a projective Calabi-Yau variety, Serre duality tells us that its derived category of coherent sheaves is a Calabi-Yau category. In our situation, however, $M$ is not projective and Serre duality does not hold. Another aspect of the lack of projectivity is that sheaves can have infinite dimensional cohomology. One definition of a Calabi-Yau algebra is that the derived category of finite dimensional modules is a Calabi-Yau category. This is \textit{not} equivalent to the derived category of $M$, and seems to often (always?) be Calabi-Yau. It would be interesting to understand this relationship further.\footnote{As the varieties in this paper are proper over affine varieties, this issue may be resolved by a proposition in \cite{Ginzburg:toapcy}.}

There are a number of physical questions that arise at this point. The first question is, what is the physical significance of the equivalence of categories that we have constructed. I do not have a good answer for this. It implies, as mentioned above, that all topological open and closed string amplitudes can be expressed in terms of the data of the quiver. This is not implausible because we have only considered the local region near the singularity. It is, in a sense, a topological version of open/closed string duality.

This equivalence of categories allows us to directly identify the fractional branes. The space-filling D3-brane can be thought of as a skyscraper sheaf located on the zero section of $M$. The representation of the quiver corresponding to this sheaf is $\mathbf{R}\Hom(\pi^*T,\mcO_p)$ with the obvious action of $B$. As the higher Exts vanish, this is an actual representation rather than a complex of representations, and the dimension of the image of the idempotent corresponding to node $a$ is $d_a = \dim \Hom(\pi^*E_a,\mcO_p) = \text{rank}(E_a)$ for all the $E_a$ locally free, \ie vector bundles. This corresponds to the gauge group $\times U(d_a)$. Recall that the fractional branes should add one to the rank of the gauge group at the corresponding node. For a node $b$, this corresponds to a `simple representation' of the quiver as we will describe below. Our equivalence of categories immediately gives us a complex of sheaves that corresponds to this representation. This is the fractional brane, $L_a$. The fractional branes turn out to be closely related to what is known as the dual collection \cite{BridgeTStruct}. It should be the case that at the point of marginal stability, $\mcO_p$ can decay to $\bigoplus_a L_a^{\oplus d_a}$. The existence of this decay is argued for in \cite{Aspinwall:2004mb}.

Next, one might ask about the significance of the existence of various different quivers. One can show \cite{Bergman:2005kv} that the moduli space of all quivers derived by these methods contain as a component the original conical singularity. Thus, in a general sense, all these theories are Seiberg dual \cite{Berenstein:2002fi}. Certain operations that take one exceptional collection to another can explicitly
be seen to be more traditional Seiberg duality \cite{Herzog:2004qw}. Nonetheless, the specific decay into fractional branes should be physical. It depends on the stability condition, however, and thus on the K\"ahler information. Bridgeland \cite{Bridgeland:2002sc} has proposed a definition of stability conditions on derived categories. Given a description of a derived category in terms of representations of a quiver, it is particularly simple to write down a stability condition on the derived category. It would be interesting to completely understand the connection between the stability condition and the information of the quiver gauge theory, in particular the values of the Fayet-Iliopoulis terms. Some connections along these lines are made in \cite{Aspinwall:2004mb}.

Finally, one might ask of the physical significance of the branes corresponding to the $\pi^*E_a$. They correspond to projective representations of the quiver. As we have seen, they generate the derived category, so perhaps it is best to think of them as a choice of basis. The construction of the fractional branes, which correspond to the simple representations of the quiver, fundamentally depended on the equivalence of categories determined by the exceptional collection.

\section{Constructing the gauge theory}\label{sec:homalg}

\subsection{The homological algebra of quivers}

Before giving the construction of the quiver gauge theory, we will need to introduce a number of facts about the homological algebra of quivers which will be used for the rest of the paper. Recall that a quiver, $Q$, is a directed graph\footnote{A nice introduction to the representation theory of quivers is \cite{Derksen}.}. We will denote the set of nodes as $\text{Nodes}(Q)$ and arrows as $\Arr(Q)$. We have two maps $s, t : \Arr(Q) \to \text{Nodes}(Q)$ giving the source and target of the arrow. A representation, $V$, of a quiver consists of a set of vector spaces\footnote{These vector spaces are often denoted $V[i]$ in the literature, but I am using parentheses to avoid confusion with the shift functor.}, $V(i)$, for every $i \in \text{Nodes}(Q)$ and maps, $V(s(a)) \to V(t(a))$, for every $a \in \Arr(Q)$. Associated to any quiver is its \textit{path algebra}, $\BC Q$, consisting of all paths in the quiver (including zero-length paths for each node) with the obvious multiplication. The source and target maps extend obviously to paths. We will denote paths composing from right to left. A $\BC Q$-module, $V$, is the same thing as a representation of the quiver. The zero length paths, $e_i$, for $i \in \text{Nodes}(Q)$ obey $e_i^2 = e_i$ and are thus idempotents. We have $V(i) = e_iV$. There are two distinguished sets of representations of $Q$. The first are the simple representations given by $S_i(j) = \MC{\delta_{ij}}$ with all maps set to zero. The second are the projective representations, $P_i \defeq \BC Q e_i$ where $P_i(j) = e_j \BC Q e_i$ is the vector space spanned by all paths from node $i$ to $j$. It is straightforward to see that $\Hom(P_i,V) \cong V(i)$. In particular, $\Hom(P_i,P_j)$ is spanned by paths from $j$ to $i$.

We will now add relations to our quiver. An admissible relation is an element of $e_i \BC Q e_j$ for some $i,j \in \text{Nodes}(Q)$. This means that it is a sum of paths with the same beginning and end. An admissible ideal is an ideal in $\BC Q$ generated by admissible relations. A \textit{quiver with relations} is given by a pair $(Q,I)$ with $I$ an admissible ideal. Its path algebra is $\BC Q / I$. A representation is either a module of this algebra, or, equivalently, a representation of $Q$ such that all maps corresponding to paths in the ideal $I$ are zero. The simple modules and projective modules are defined similarly as in the case without relations and satisfy the same properties.

Let $S$ denote the direct sum of the simple representations. Let $J$ denote the ideal generated by all paths of length greater than or equal to one. Then, we have the following canonical maps (taken from \cite{Keller:2000ai}):
\begin{eqnarray}
\label{extarr}
&\bigoplus_{i=1}^\infty \Ext^1(S,S)^{\otimes i}\to J^\vee \ ,& \\
\label{extrels}
&\Ext^2(S,S) {\buildrel\sim\over\to} (I / (IJ + JI))^\vee &
\end{eqnarray}
where the dual is as a $\BC$-vector space. Both of these are related to existence of a canonical projective resolution for any representation of the quiver with relations. This is related to a resolution of the path algebra $\BC Q / I$ in the category of bimodules \cite{Bardzell:1997as} which I will now describe.

To save space, let us denote $A \defeq \BC Q / I$. We choose a minimal basis of admissible relations generating $I$. The set of all such relations will be denoted $\text{Rel}(I)$. We can define source and target maps for the each relation. From \eqref{extrels}, we have $\#\{R \in \text{Rel}(I) \text { s.t. } s(R) = i \text{ and } t(R) = j\} =  \dim \Ext^2(S_i,S_j)$.

Next, let $P_{ij} = A e_i \otimes e_j A$ be a bimodule for all $i,j \in \text{Nodes}(Q)$. These are projective bimodules and will be the objects in our resolution. Given an arrow $a \in \Arr(Q)$, we have an operation $\partial_a : \BC Q \to P_{t(a)s(a)}$ defined as follows. Any element in $\BC Q$ is a sum of paths in the quiver. For each element in the sum, locate all occurrences of the arrow $a$. Then, the path can be written as $p a q$ with $t(q) = s(a)$ and $t(a) = s(p)$. This defines an element $p \otimes q \in P_{t(a)s(a)}$. If $a$ occurs multiple times in the path, then do this for each occurrence. The value of $\partial_a$ in $P_{t(a)s(a)}$ is the sum of all elements so obtained. Note that this operation is defined on $\BC Q$, \textit{not} $A$.

We now have the following resolution:
\begin{equation}
\label{bires1}
\cdots \longrightarrow \bigoplus_{R \in \text{rel}(I)} P_{t(R)s(R)} {\buildrel g\over\longrightarrow}
\bigoplus_{a \in \Arr(Q)} P_{t(a)s(a)} {\buildrel f\over\longrightarrow} \bigoplus_{i\in \text{Nodes}(Q)} P_{ii} {\buildrel m\over\longrightarrow} A \longrightarrow 0\ .
\end{equation}
The maps are as follows. $m$ is just the multiplication map on $A$. It is straightforward to see that any map from $P_{ij}$ to another bimodule is completely determined by the image of the element $e_i \otimes e_j$. We define a map $f_a : P_{t(a)s(a)} \to \bigoplus_{i} P_{ii}$ that takes $e_{t(a)}\otimes e_{s(a)} \mapsto a \otimes e_{s(a)} - e_{t(a)} \otimes a$. We then have $f = \sum_a f_a$. Finally, we define a map $g_R : P_{t(R)s(R)} \to \bigoplus_a P_{t(a)s(a)}$ by $e_{t(R)} \otimes e_{s(R)} \mapsto \sum_a \partial_a R$. These add to give the map $g$.

Later, we will assume that the global dimension of $A$ is two which allows us to replace the dots in \eqref{bires1} with a zero. This follows from the minimality of the resolution \cite{Butler:1999mr}.

Since the operation $\otimes_A^\mathbf{L} A$ is the identity, we see that the resolution \eqref{bires1} gives the first few terms of a canonical projective resolution of any module. If we apply this to the simple representation, $S_i$, we obtain the following resolution:
\begin{equation}
\label{simpres}
\cdots \longrightarrow \bigoplus_{\substack{R\in\text{Rel}(I) \\ s(R) = i}} P_{t(R)} \longrightarrow \bigoplus_{\substack{a\in\Arr(Q) \\ s(a) = i}} P_{t(a)} \longrightarrow P_i \longrightarrow S_i \longrightarrow 0\ .
\end{equation}
Applying the functor $\Hom(-,S_j)$ to this resolution, we can understand the identifications \eqref{extarr} and \eqref{extrels}. In particular, $\dim \Ext^1(S_i,S_j)$ is the number of arrows (\textit{not} paths) between nodes $i$ and $j$, and $\dim \Ext^2(S_i,S_j)$ is the number of relations.

\subsection{The quiver gauge theory}\label{subsec:qgt}

The relation between these quiver algebras and the physics of quiver gauge theories is that the matter content in the gauge theory should be given by the string states between the fractional branes. In the case that the gradings align in the topological theory, the massless chiral multiplets between two branes are given by the first Ext group between them. As we have seen, because the fractional branes correspond to the simple representations, these states are precisely the arrows in the quiver. The vector multiplets for each node correspond to the $\Ext^0(S_i,S_i)$. All other Ext groups correspond to states of string scale mass.

We will apply this to determine the quiver corresponding to $\End(\pi^*T)$. In particular, let $s : N \to M$ be the zero section of the projection $\pi$. Then, the simple representations correspond to the objects $s_*(S_i)$ in $\mcD(M)$. We would like to compute $\Ext^1(s_*S,s_*S)$. This can be done as follows. First, we apply the adjunction between pushforwards and pullbacks:
\begin{equation}
\label{pushedext}
\Hom_{\mcD(M)}(s_*S_i,s_*S_j[n]) \cong \Hom_{\mcD(N)}(\mathbf{L}s^*s_*S_i,S_j[n]) \ .
\end{equation}
Now, to compute $\mathbf{L}s^*$, we must resolve $s_*S_i \cong \pi^*S_i \otimes s_*\mcO_N$ in terms of flat (in our case, locally free) sheaves. We have the following resolution:
\begin{equation}
\label{zerores}
0\longrightarrow \pi^*\omega^{-1}_N \longrightarrow \mcO_M \longrightarrow s_*\mcO_N \longrightarrow 0\ .
\end{equation}
Assuming that tensoring with the $S_i$ is exact,\footnote{This assumption seems to be implicit in other papers on the subject.} we can tensor with $\pi^*S_i$ and apply $s^*$ giving that $\mathbf{L}s^*s_*S_i$ is quasiisomorphic to 
\begin{equation}
(\omega^{-1}_N \otimes S_i)[1] \oplus S_i\ .
\end{equation}
Substituting into \eqref{pushedext}, we obtain
\begin{equation}
\begin{split}
\label{pushextrel}
\Hom_M(s_*S_i,s_*S_j[n]) &\cong \Hom_N(S_i,S_j\otimes \omega_N[n-1]) \oplus \Hom_N(S_i,S_j[n]) \\
&\cong \Hom_N(S_j,S_i[3-n])^\vee \oplus \Hom_N(S_i,S_j[n])
\end{split}
\end{equation}
where we have used Serre duality on $N$ in the last equality.

Thus, we see that in the quiver gauge theory, we have bifundamental matter for every arrow in the quiver for $N$ and additional arrows in the opposite direction of every relation in $\text{Rel}(I)$. Note that an $\Ext^0$ between different branes gives a tachyonic state, so we have to require that there are no $\Ext^3$s or higher between any of the simples. This is is equivalent to the statement that the global dimension of $A$ is 2.

Let us denote by $\barQ$ the quiver so obtained. We would like to know the set of relations such that $\BC \barQ / \bar{I} \cong \End(\pi^*T)$. We see from \eqref{pushextrel} and \eqref{extrels} that there is a relation on $\barQ$ for every relation for $\End(T)$ and a relation in the opposite direction for every arrow in $Q$. This leads to a natural conjecture. Choose a basis of relations $R_a$ that generate $I$ as above. The quiver $\barQ$ has all the arrows of $Q$ plus, for each $R_a$, an additional arrow which we will denote $r_a$ and which satisfies $s(r_a) = t(R_a)$ and $t(r_a) = s(R_a)$. 

The vector space $L \defeq \BC \barQ / [\BC \barQ, \BC \barQ]$ consists of all loops in the quiver. Thus, we can define an object $W$ consisting of the sum of the loops $\sum_{a}r_aR_a$. This is called the superpotential of the theory. We can define two operations from $L \to \BC \barQ$. The first $\iota_n : L \to \Hom(P_n,P_n)$ turns the loop into a path based on the node $n$ (any elements that do not go through $n$ go to zero). The second $\del_a : L \to \BC \barQ$ takes any occurrences of the arrow $a$ in the loop and removes them leading to a path from the target of $a$ to its source. This can be thought of, formally, as $\del_a(\ell) = \iota_{s(a)}(\ell) a^{-1}$. This should not be confused with the other $\del_a$ defined above, but we will make use of this abuse of notation later.

Let $\bar{I}$ be the ideal generated by all paths of the form $\del_a W$ for all arrows $a \in \Arr(\barQ)$. The claim is that $B = \End(\pi^*T) \cong \BC \barQ/ \bar{I}$. This is argued for in \cite{Aspinwall:2005ur} by examining the A$_\infty$-algebra of the Exts of the simples, and has been proven in \cite{Segal:2007ad}.

From the physics point of view, $\barQ$ and $W$ determine a quiver gauge theory. The ranks of the gauge groups are determined as above by $d_i = \Hom_M(\pi^*E_i,\mcO_p) = \text{rank}(E_i)$. The values of the Fayet-Ilioupolis term depend on K\"ahler information and are, as such, not visible in this construction. As mentioned above, this construction has survived many nontrivial consistency checks, but there is still an element of conjecture to it.

\section{Deformations of the quiver gauge theory}\label{sec:deform}

\subsection{The self-dual resolution}\label{ssec:sdres}

In this section, we will relate superpotential deformations of the quiver gauge theory as constructed above. We will now \textit{assume} that the algebra, $A = \BC \barQ / \bar{I}$ is a Calabi-Yau algebra. This will allow us to take advantage of an extension of the projective resolution \eqref{bires1} \cite{Bocklandt:2006gc,Ginzburg:toapcy}.
\begin{equation}
\label{bires2}
0\longrightarrow \bigoplus_{i\in \text{Nodes}(\barQ)} P_{ii} {\buildrel h\over\longrightarrow} \bigoplus_{a \in \Arr(\barQ)} P_{s(a)t(a)} {\buildrel g\over\longrightarrow}
\bigoplus_{a \in \Arr(\barQ)} P_{t(a)s(a)} {\buildrel f\over\longrightarrow} \bigoplus_{i\in \text{Nodes}(\barQ)} P_{ii} {\buildrel m\over\longrightarrow} A \longrightarrow 0\ .
\end{equation}
Given the pairing of arrows with relations, we have turned the sum in the second term into a sum over arrows rather than relations. The maps $m$, $f$ and $g$ are the same as above. The map $h$ is given by $e_i \otimes e_i \mapsto e_i \otimes a - a \otimes e_i$. The exactness of the sequence is proven in the references. It also obeys a nice self-duality property relating to the Calabi-Yau structure.

We now wish to compute the Hochschild cohomology using this resolution. By definition, to do this we apply the functor $\Hom_{A-A}(-,A)$ to the above and take the cohomology of the resulting sequence. Using what we know about maps from the bimodules $P_{ij}$, we see that $\Hom_{A-A}(P_{ij},A) \cong \Hom_A(P_i,P_j)$ or paths from node $j$ to node $i$. The maps are straightforward to determine.

Concentrating on $\HH^2(A)$, we wish to determine closed elements in the vector space $\bigoplus_a \Hom_A(P_{s(a)},P_{t(a)})$. The map to $\bigoplus_i \Hom_A(P_i,P_i)$ is given by $p \in \Hom_A(P_{s(a)},P_{t(a)}) \mapsto pa - ap$. Pick an $\ell \in L$ where $L \defeq \BC \barQ / [\BC \barQ, \BC \barQ]$ as above. Then we have $\sum \del_a\ell \in \Hom_A(P_{s(a)},P_{t(a)})$ where we have composed with the projection from $\BC \barQ \to A$. It is straightforward to verify that this is closed. These will be the superpotential deformations.

It is interesting to ask about the exact elements. We can now form the element $\del^2_{ba} W \defeq \del_b \del_a W \in P_{t(b)s(b)}$. Given an element $q \in \Hom_{A-A}(P_{t(b)s(b)}, A)$, we apply it to $\del^2_{ba}W$. One can see immediately that the result obeys $s(q(\del^2_{ba}W)) = t(a)$ and $t(q(\del^2_{ba}W)) = s(a)$, so it lives in $\Hom_A(P_{s(a)},P_{t(a)})$. The exact elements are then the sums $\sum_a q(\del^2_{ba}W)$. Note that $\del_a W$ is exact.

\subsection{Relation to the bar resolution}

We will now show that the element $\sum \del_a \ell$ does correspond to a superpotential deformation. We will do this by exhibiting the beginning of a quasiisomorphism between the resolution \eqref{bires2} and the bar resolution from \eqref{barres}:
\begin{equation}
\label{barquasi}
\begin{split}
\begindc{\commdiag}[5]
	\obj(10,10)[db]{$\cdots$}
	\obj(30,10)[Pr]{$\bigoplus P_{s(a)t(a)}$}
	\obj(50,10)[Pa]{$\bigoplus P_{t(a)s(a)}$}
	\obj(70,10)[Pn]{$\bigoplus P_{ii}$}
	\obj(90,10)[Ab]{$A$}
	\obj(10,20)[dt]{$\cdots$}
	\obj(30,20)[AAAA]{$A \otimes A \otimes A \otimes A$}
	\obj(50,20)[AAA]{$A \otimes A \otimes A$}
	\obj(70,20)[AA]{$A \otimes A$}
	\obj(90,20)[At]{$A$}
	\mor{db}{Pr}{}
	\mor{Pr}{Pa}{}
	\mor{Pa}{Pn}{}
	\mor{Pn}{Ab}{}
	\mor{dt}{AAAA}{}
	\mor{AAAA}{AAA}{}
	\mor{AAA}{AA}{}
	\mor{AA}{At}{}
	\mor{AAAA}{Pr}{$f$}
	\mor{AAA}{Pa}{$g$}
	\mor{AA}{Pn}{$h$}
\enddc
\end{split}\ .
\end{equation}
We will use $\delta$ to denote the horizontal morphisms in the bar complex and $d$ for the resolution \eqref{bires2}. We will also make the choice of a map $\sigma : \BC\barQ / \bar{I} \to \BC\barQ$ that is a section of the projection. It is necessarily a map of vector spaces and not a map of algebras. The map $h$ is specified by $1 \mapsto \bigoplus e_i \otimes e_i$. The map $g$ is specified by $b \in A \mapsto \sum_\alpha \del_\alpha \sigma(b)$. For the element $[1|b|1] \in A \otimes A \otimes A$, we have $h\delta[1|b|1] = h([b|1] - [1|b]) = b \otimes e_{s(b)} - e_{t(b)} \otimes b$ and $dg([1|b|1]) = d(\sum_\alpha \del_\alpha \sigma(b)) = b \otimes e_{s(b)} - e_{t(b)} \otimes b$ where the latter equality comes from a telescoping sum.

Finally, for the map $f$, we need the image of $[1|b|c|1]$. We have 
\begin{equation}
\begin{split}
g\delta[1|b|c|1] &= g([b|c|1] - [1|bc|1] + [1|b|c] = \bigoplus_\alpha b\del_\alpha \sigma(c) - \del_\alpha \sigma(bc) + (\del_\alpha \sigma(b)) c \\
 &= \bigoplus_\alpha \del_\alpha(\sigma(b)\sigma(c) - \sigma(bc))\ .
 \end{split}
 \end{equation}
We define $\tau(b,c) \defeq \sigma(b)\sigma(c) - \sigma(bc) \in \bar{I}$. 

Let $\widetilde{P}_{ij}$ be the analog of the $P_{ij}$ for the algebra $\BC \barQ$. Then, we have a map $j : \widetilde{P}_{s(a)t(a)} \to \bar{I}$ which takes $e_{s(a)} \otimes e_{t(a)} \mapsto R_a$. This is a surjective map of bimodules of $\BC \barQ$, and considered as a map of vector spaces we can choose a section. Composing with the projection $\widetilde{P}_{s(a)t(a)} \to P_{s(a)t(a)}$, we obtain a map $\varphi : \bar{I} \to P_{s(a)t(a)}$. Finally, define $f([1|b|c|1]) = \varphi(\tau(b,c))$. We have that $df = g\delta$ by construction.

Recall, now, that the putative superpotential deformation comes from an element $\ell \in L$. This defines a map $\Lambda : \bigoplus P_{s(a)t(a)} \to A$ by $e_{s(a)}\otimes e_{t(a)} \mapsto \del_a \ell$. We can compose this with the map $f$ to obtain a map from $A^{\otimes 4} \to A$. Recall from \eqref{defprod}, that the image of $[1|b|c|1]$ is the first order deformation of the product. Thus, we have $\mu(b,c) = bc + \hbar \Lambda(\varphi(\tau(b,c))) + \mathscr{O}(\hbar^2)$.

To get an intuition for this map, we observe that $\tau(b,c)$ is a formalization of the idea of when a product $bc$ is ``part of a relation". In particular, if $\sum_i \sigma(b_i) \sigma(c_i) \in \bar{I}$, then it is equal to $\sum_i \tau(b_i,c_i)$. We can think of the map $\varphi$ as taking an element in $\bar{I}$, expressing it in terms of the relations and giving the following:
\begin{equation}
\sum_i \epsilon_i R_i \rho_i \mapsto \sum \epsilon_i \otimes \rho_i\ ,
\end{equation}
and
\begin{equation}
\Lambda(f([1|b|c|1])) = \Lambda\big(\varphi(\tau(b,c))\big) = \sum_i \epsilon_i (\del_i\ell) \rho_i\ .
\end{equation}
Unpacking this, we see that, whenever $bc$ is ``part of a relation", we take that part of the relation and replace it with a derivative of $\ell$. With a little thought, one can see that this exactly implements the addition of $-\hbar \ell$ to the superpotential. For example, note that if we have $\sum_i \sigma(b_i) \sigma(c_i) = \del_a W$ for some $a \in \Arr(\barQ)$,
\begin{equation}
\begin{split}
\sum_i \mu(b_i,c_i)&= \sum_i b_ic_i + \hbar \Lambda(\varphi(\tau(b,c))) + \mathscr{O}(\hbar^2) \\
	&= 0 + \hbar \Lambda\left(\varphi\left(\sum_i \sigma(b_i) \sigma(c_i)\right)\right) + \mathscr{O}(\hbar^2) \\
	&= \hbar \Lambda(\varphi(R_a)) + \mathscr{O}(\hbar^2) \\
	& = \hbar \Lambda\left(e_{t(a)} \otimes e_{s(a)}\right) + \mathscr{O}(\hbar^2) \\
	&= \hbar \del_a \ell + \mathscr{O}(\hbar^2) \ .
\end{split}
\end{equation}
This corresponds to a deformation of the relation $R_a \leadsto R_a - \hbar \del_a \ell = \del_a(W - \hbar \ell)$. This concludes the proof that the element of $\HH^2(A)$ that we have identified corresponds to a superpotential deformation.

\section{Noncommutative deformations}\label{sec:ncdef}

\subsection{From bimodules to Fourier-Mukai transforms}\label{subsec:bimod}

In this section, we will turn the bimodule resolution \eqref{bires2} into an object in $\mcD(M \times M)$. Because \eqref{bires2} implements the identity functor, the object in $\mcD(M \times M)$ so obtained must be isomorphic to the diagonal $\Delta_* \mcO_M$. This will allow us to relate the Hochschild cohomology element computed in section \ref{ssec:sdres} to an element in sheaf cohomology. In principle, this provides an explicit map between deformations of the quiver gauge theory and deformations of the geometry (the sheaf cohomology groups) using the quasiisomorphism \eqref{HKRiso} from section \ref{ssec:geo}:
\begin{equation}
\label{HKRiso2}
I_{\mathrm{HKR}} : \bigoplus_{i=0}^{\dim X} \wedge^i TX [-i] {\buildrel\sim\over\longrightarrow} \mathscr{E}xt_{X \times X}(\mcO_\Delta,\mcO_\Delta)\ .
\end{equation}
As mentioned above, the Atiyah-Chern character gives this quasiisomorphism, but it can be a bit difficult to implement in practice. It is straightforward to compute that the noncommutative deformations are given by $\wdg^2 TM = \pi^*\omega^{-1}_N \oplus \pi^*(TN \otimes \omega_N)$. In the case that $H^1(TN \otimes \omega_N^{-p})$ vanishes for all $p > 0$, we will demonstrate an explicit map between these noncommutative deformations and the corresponding superpotential deformations, avoiding the use of the Atiyah-Chern characters.

We have a number of functors floating around. Recall that $B = \End_{\mcD(M)}(\pi^*T)^\text{op}$. We then have the functors $\mathbf{R}\Hom_{\mcD(M)}(\pi^*T, -) : \mcD(M) \to \mcD(B \mathrm{- mod})$ and $- \otimes^\mathbf{L}_B \pi^*T : \mcD(B \mathrm{- mod}) \to \mcD(M)$ whose composition is equivalent to the identity. We also have a functor given by an object $X \in \Obj(\mcD(B - B))$, $X \otimes^\mathbf{L}_B - : \mcD(B \mathrm{- mod}) \to \mcD(B \mathrm{- mod})$. We can compose these three functors to obtain a functor from $\mcD(M) \to \mcD(M)$. By the theorem of To\"en \cite{Toen:2004to}, this corresponds to an element in $\mcD(M \times M)$. (Technically, this should be the derived category of quasicoherent sheaves, but we will see that everything is nicely coherent and our representative will consist of bounded complexes of locally free sheaves.) We will apply this procedure to the complex \eqref{bires2} which is quasiisomorphic to $B$, \ie the identity functor. This must, then, give rise to something isomorphic to the diagonal $\mcO_\Delta = \Delta_* \mcO_M$ in $\mcD(M \times M)$.

As a first step, let us find the object in $\mcD(M \times M)$ corresponding to the bimodule $P_{ij}$. We will apply the three functors to a test object $\mcF$:
\begin{equation}
\begin{split}
\mcF  &\mapsto \mathbf{R}\Hom_{\mcD(M)}(\pi^*T, \mcF)\ ,\\
           &\mapsto Be_i \otimes e_j B \otimes^\mathbf{L}_B  \mathbf{R}\Hom_{\mcD(M)}(\pi^*T, \mcF)\ ,\\
           &\mapsto \left(\pi^*T \otimes^\mathbf{L}_B Be_i\right) \otimes \left(e_j B \otimes^\mathbf{L}_B
           	  \mathbf{R}\Hom_{\mcD(M)}(\pi^*T, \mcF)\right)\ .
\end{split}
\end{equation}
Note that the object on the left is a complex of sheaves, while the object on the right is a complex of vector spaces. Recall that $\pi^*T = \bigoplus_n \pi^*E_n$ and that $e_n \in B$ corresponds to the identity element in $\End_{\mcD(M)}(\pi^*E_n)$. The fact that $e_m e_n = 0$ for all $m\neq n$ implies that
\begin{equation}
\begin{split}
\pi^*T \otimes^\mathbf{L}_B Be_i &= \pi^*E_i \ , \\
	e_j B \otimes^\mathbf{L}_B \mathbf{R}\Hom_{\mcD(M)}(\pi^*T, \mcF) &=
	\mathbf{R}\Hom_{\mcD(M)}(\pi^*E_j,\mcF)\ ,
\end{split}
\end{equation}
and, thus,
\begin{equation}
\label{eiejmap}
\mcF \mapsto \pi^*E_i \otimes \mathbf{R}\Hom_{\mcD(M)}(\pi^*E_j,\mcF)\ .
\end{equation}

Recall from \eqref{FMT} that for an object $\mcM$, the Fourier-Mukai transform with kernel $\mcM$ of an object $\mcF$ is given by:
\begin{equation}
\label{FMT2}
\mcF \mapsto \mathbf{R}\pi_{1*}(\mcM \otimes^\mathbf{L} \pi_2^*(\mcF))\ .
\end{equation}
It is straightforward to verify that the map \eqref{eiejmap} is precisely implemented as a Fourier-Mukai transform with kernel $\mcP_{ij} \defeq \pi^*E_i \boxtimes (\pi^*E_j)^\vee$. Here, $\mcE^\vee \defeq \mathscr{H}om(\mcE,\mcO)$ as we will only be dealing with locally free sheaves.

Because everything in sight is functorial, we can translate \eqref{bires2} into the following complex in $\mcD(M \times M)$ necessarily isomorphic to $\mcO_\Delta$:
\begin{equation}
\label{sbires2}
0\longrightarrow \bigoplus_{i\in \text{Nodes}(\barQ)} \mcP_{ii} \longrightarrow \bigoplus_{a \in \Arr(\barQ)} \mcP_{s(a)t(a)}\longrightarrow
\bigoplus_{a \in \Arr(\barQ)} \mcP_{t(a)s(a)}\longrightarrow \bigoplus_{i\in \text{Nodes}(\barQ)} \mcP_{ii}  \longrightarrow 0\ .
\end{equation}

We also have a complex in $\mcD(N \times N)$ coming from \eqref{bires1}. Abusing the notation to let $\mcP$ denote the relevant sheaf on $N \times N$, we have:
\begin{equation}
\label{sbires1}
0\longrightarrow  \hspace{-.7cm}\bigoplus_{a \in (\Arr(\barQ)-\Arr(Q))}\hspace{-.7cm} \mcP_{s(a)t(a)}\longrightarrow
\bigoplus_{a \in \Arr(Q)} \mcP_{t(a)s(a)}\longrightarrow \bigoplus_{i\in \text{Nodes}(Q)} \mcP_{ii}  \longrightarrow 0\ .
\end{equation}
By the same reasoning as above, this is isomorphic to the structure sheaf of the diagonal in $N \times N$ which we will also denote, $\mcO_\Delta$.

We will now apply the functor $\mathbf{R}\mathscr{H}om_{M}(\mathbf{L}\Delta^*(-),\mcO_M)$ to both complexes. We introduce the following notation to save space:
\begin{equation}
\begin{split}
\mcO(i) = \mathscr{H}om_N(E_i,E_i)  \qquad &\text{for } i \in \text{Nodes}(\barQ)\ , \\
\mcO(a) = \mathscr{H}om_N(E_{t(a)},E_{s(a)}) \qquad &\text{for } a \in \Arr(\barQ)\ , \\
\mcO(-a) = \mathscr{H}om_N(E_{s(a)},E_{t(a)}) \qquad &\text{for } a \in \Arr(\barQ)\ .
\end{split}
\end{equation}
Then, $\mathbf{R}\mathscr{H}om_{M }(\mathbf{L}\Delta^*\mcO_\Delta,\mcO_M)$ is isomorphic to
\begin{equation}
\label{diagcomM}
0  \longrightarrow \bigoplus_{i\in \text{Nodes}(\barQ)} \pi^*\mcO(i) 
\longrightarrow \bigoplus_{a \in \Arr(\barQ)} \pi^*\mcO(a) 
\longrightarrow \bigoplus_{a \in \Arr(\barQ)} \pi^*\mcO(-a)
\longrightarrow \bigoplus_{i\in \text{Nodes}(\barQ)} \pi^*\mcO(i)
\longrightarrow 0\ .
\end{equation}
Because the $E_i$ form a strong exceptional collection, all the objects in the above complex are acyclic, and we can use this complex to compute $\Ext^i_M(\mathbf{L}\Delta^*\mcO_\Delta,\mcO_M) \cong \Ext^i_{M \times M}(\mcO_\Delta,\mcO_\Delta) = \HH^i(M)$ as in \eqref{HHspacedef}. This precisely corresponds to the calculation of section \ref{ssec:sdres}. If we now were to compute the quasiisomorphism to the sheaf cohomology of this complex, we could then explicitly see the correspondence between superpotential deformations and geometrical deformations given the HKR isomorphism.

%\begin{equation}
%\begin{split}
%0 & \longrightarrow \bigoplus_{i\in \text{Nodes}(\barQ)} \mathscr{H}om(\pi^*E_i,\pi^*E_i) 
%\longrightarrow \bigoplus_{a \in \Arr(\barQ)} \mathscr{H}om(\pi^*E_{t(a)},\pi^*E_{s(a)}) \\
%& \qquad \longrightarrow \bigoplus_{a \in \Arr(\barQ)}\mathscr{H}om(\pi^*E_{s(a)},\pi^*E_{t(a)})
%\longrightarrow \bigoplus_{i\in \text{Nodes}(\barQ)} \mathscr{H}om(\pi^*E_i,\pi^*E_i) 
%\longrightarrow 0\ .
%\end{split}
%\end{equation}

In what follows, we will demonstrate the explicit embedding for noncommutative deformations in many cases. In order to do so, we will need to understand the HKR map for $N$.  By applying the same procedure as above for \eqref{sbires1}, we obtain that
$\mathbf{R}\mathscr{H}om_{N}(\Delta^*\mcO_\Delta,\mcO_M)$ is isomorphic to
\begin{equation}
\label{diagcomN}
0 \longrightarrow \bigoplus_{i\in\text{Nodes}(Q)} \mcO(i) 
\longrightarrow \bigoplus_{a \in \Arr(Q)} \mcO(a)
\longrightarrow \hspace{-0.7cm}\bigoplus_{a \in (\Arr(\barQ) - \Arr(Q))}\hspace{-0.7cm} \mcO(-a) \longrightarrow 0
\end{equation}

\subsection{Finding the anticanonical sheaf}

For our next step, we would like to get a handle on the anticanonical sheaf. Let our exceptional collection be labelled $E_0 \dots E_n$ and the simple objects be $S_0 \dots S_n$. Because $\Hom(E_n,E_i) = 0$ for all $i \neq n$, it follows that $E_n$ is the projective for an initial node in the quiver, \ie there are no arrows with $t(a) = n$. It is an elementary consequence of the construction of the simples in terms of mutation \cite{BondalQuiv} that $\mathscr{H}om(S_n,E_n) \cong \omega_N^{-1}[-2]$. Now, recall the resolution \eqref{simpres}. Translated into sheaves, we obtain that $S_n$ is isomorphic to:
\begin{equation}
\label{simpres2}
0 \longrightarrow \bigoplus_{R} E_{t(R)} \longrightarrow \bigoplus_{\substack{a \in \Arr(Q)\\ s(a) = n}} E_{t(a)} \longrightarrow E_n  \longrightarrow 0\ .
\end{equation}
We apply the functor $\mathscr{H}om(-,E_n)$ to this resolution to obtain:
\begin{equation}
\label{initres}
0 \longrightarrow \mcO(n) \longrightarrow \bigoplus_{\substack{a \in \Arr(Q)\\ s(a) = n}} \mcO(a) \longrightarrow \hspace{-0.7cm}\bigoplus_{\substack{b \in (\Arr(\barQ) - \Arr(Q))\\  t(b) = n}}\hspace{-0.7cm} \mcO(-b) \longrightarrow 0\ .
\end{equation}
The map between the last two terms is given by composition with $\del_a R_b$ which in this case can be thought of as an element in $\Hom(E_{t(R_b)},E_{t(a)})$ as there are no loops in the quiver $Q$.
The isomorphism with $\omega_n^{-1}[-2]$ tells us that there is an isomorphism 
\begin{equation}
\label{acmap}
\omega_N^{-1} {\buildrel\sim\over\longrightarrow} \text{Coker}(\del_a R_b) =  \bigoplus \mcO(-b) \bigg/ \text{Im } \del_a R_b \ .
\end{equation}
As $H^1(\pi^*\mcO(n)) = 0$ by exceptionality, we have:
\begin{equation}
\label{anticonmap}
H^0(\omega_N^{-1}){\buildrel\sim\over\longrightarrow} H^0\left(\bigoplus \mcO(-b)\right) \bigg/ \text{Im } \del_a R_b\ .
\end{equation}
Unpacking this notation, we have a map of sections of $\omega_N^{-1}$ to a quotient of paths from the initial node in the quiver, $n$, to the targets of all relations beginning at $n$.

Now recall the construction of the quiver $\barQ$. From \eqref{loopshom}, we have that the space of loops in the path algebra based at $n$ is given by $\Hom_N(E_n,E_n \otimes \omega^{-1}_N) = H^0(\mathscr{H}om(E_n,E_n) \otimes \omega^{-1}_N)$. Tensoring with the identity section of $\mathscr{H}om(E_n,E_n)$ gives us a map from $H^0(\omega^{-1}_N)$ into the space of loops based at $n$. Pick a representative of this in $\BC \barQ$. This is a sum of loops each of which starts at node $n$, travels along the arrows of $Q$ and eventually returns to $n$ by way of an arrow in $\barQ$ but not in $Q$ (it cannot go `backwards' twice because of the grading in \eqref{loopshom}). The choice of a representative in $\BC \barQ$ is arbitrary up to an element of $\bar{I}$. As we are looking at $\BC Q / I$, we only need to worry about relations that contain arrows in $\Arr(\barQ) - \Arr(Q)$. It is then straightforward to see that this ambiguity is exactly the image of $\del_a R_b$.

Finally, due to the ampleness of $\omega^{-1}_N$ and the flatness of $\pi$, all of this pulls back to a map
\begin{equation}
\label{anticonmapM}
H^0(\pi^*\omega_N^{-1}){\buildrel\sim\over\longrightarrow} H^0\left(\bigoplus_{\substack{b \in (\Arr(\barQ) - \Arr(Q))\\  t(b) = n}} \hspace{-0.7cm}\pi^*\mcO(-b)\right) \Bigg/ \text{Im } \del_a R_b\ .
\end{equation}

\subsection{Hochschild cohomology for $N$}

The HKR theorem on $N$ tells us that $\HH^2(N) \supset H^0(\omega^{-1}_N)$. The goal of this section is to find an explicit map from $H^0(\omega^{-1}_N)$ as presented above into $\HH^2(N)$. We have the following commutative diagram:
\begin{equation}
\label{omegaembed}
\begin{split}
\begindc{\commdiag}[3]
	\obj(10,40)[tlz]{$0$}
	\obj(40,40)[nodes]{$\displaystyle \bigoplus_{i\in\text{Nodes}(Q)} \mcO(i)$}
	\obj(80,40)[arrows]{$\displaystyle \bigoplus_{a \in \Arr(Q)} \!\!\!\mcO(a)$}
	\obj(120,40)[rels]{$\displaystyle \bigoplus_{b \in (\Arr(\barQ) - \Arr(Q))}\hspace{-0.8cm} \mcO(-b)$}
	\obj(150,40)[trz]{$0$}
	\obj(10,10)[blz]{$0$}
	\obj(40,10)[bnode]{$\mcO(n)$}
	\obj(80,10)[barrows]{$\displaystyle \bigoplus_{\substack{a \in \Arr(Q)\\ s(a) = n}} \!\!\!\mcO(a)$}
	\obj(120,10)[brels]{$\displaystyle \bigoplus_{\substack{b \in (\Arr(\barQ) - \Arr(Q))\\  t(b) = n}}\hspace{-0.8cm} \mcO(-b)$}
	\obj(150,10)[brz]{$0$}
	\mor{tlz}{nodes}{}
	\mor{nodes}{arrows}{}
	\mor{arrows}{rels}{}
	\mor{rels}{trz}{}
	\mor{blz}{bnode}{}
	\mor{bnode}{barrows}{}
	\mor{barrows}{brels}{}
	\mor{brels}{brz}{}
	\mor(40,10)(40,37){}[\atright,\injectionarrow]
	\mor(80,15)(80,37){}[\atright,\injectionarrow]
	\mor(120,15)(120,37){}[\atright,\injectionarrow]
\enddc
\end{split}\ .
\end{equation}
The commutativity of this diagram follows from the fact that $n$ is an initial node in the quiver. Again, every sheaf in sight is acyclic by the exceptionality of the $E_i$. After taking global sections, the  cohomology of the top row is $\HH^*(N)$ and the cohomology of the bottom row is $H^0(\omega_N^{-1})$. Composing with the rightmost arrow of the diagram, the map \eqref{anticonmap} gives us the embedding of $H^0(\omega_N^{-1})$ into $\HH^2(N)$.

\subsection{Hochschild cohomology for $M$}

It is an easy calculation to see that 
\begin{equation}
\label{HHsplit}
\bigoplus_{i=0}^3 \wdg^i TM [-i] \cong \bigoplus_{i=0}^2 \big(\wdg^i\pi^*TN[-i] \oplus \pi^*\Omega_N^i[i-3]\big)\ .
\end{equation}
We will now find this decomposition in the complex \eqref{diagcomM}. We begin with the following commutative diagram. To save space,  we will implicitly sum over the following sets when the relevant letter occurs: $i \in\text{Nodes}(\barQ)$, $a \in \Arr(Q)$ and $A \in \Arr(\barQ) - \Arr(Q)$.
\begin{equation}
\label{NtoM}
\begin{split}
\begindc{\commdiag}[5]
	\obj(20,10)[blz]{$0$}
	\obj(30,10)[bnodes]{$\displaystyle \pi^*\mcO(i)$}
	\obj(50,10)[barr]{$\displaystyle \pi^*\mcO(a)$}
	\obj(75,10)[brel]{$\displaystyle \pi^*\mcO(-A)$}
	\obj(95,10)[brz]{$0$}
	\obj(105,10)[brrz]{$0$}
	\obj(20,20)[mlz]{$0$}
	\obj(30,20)[mnodes]{$\displaystyle  \pi^*\mcO(i)$}
	\obj(50,20)[marr]{$\displaystyle  \pi^*\mcO(a) \oplus \pi^*\mcO(A)$}
	\obj(75,20)[mrel]{$\displaystyle \pi^*\mcO(-A) \oplus \pi^*\mcO(-a)$}
	\obj(95,20)[mnodes2]{$\displaystyle \pi^*\mcO(i)$}
	\obj(105,20)[mrz]{$0$}
	\mor{blz}{bnodes}{}
	\mor{bnodes}{barr}{}
	\mor{barr}{brel}{}
	\mor{brel}{brz}{}
	\mor{brz}{brrz}{}
	\mor{mlz}{mnodes}{}
	\mor{mnodes}{marr}{}
	\mor{marr}{mrel}{}
	\mor{mrel}{mnodes2}{}
	\mor{mnodes2}{mrz}{}
	\mor{mnodes2}{brz}{}
	\mor{mrel}{brel}{}
	\mor{marr}{barr}{}
	\mor{mnodes}{bnodes}{}
\enddc
\end{split}\ .
\end{equation}
The top line is \eqref{diagcomM} and the bottom line is the pullback of \eqref{diagcomN}. All the vertical arrows are the projections onto components. Commutativity is easy to verify. The complex \eqref{diagcomM} has the remarkable property that taking its dual gives exactly the same complex. This is related to the Calabi-Yau property of the algebra \cite{Bocklandt:2006gc}.

By taking the dual of this diagram, we obtain another commutative diagram as everything is locally free. This gives a chain map from the dual of \eqref{diagcomN} to \eqref{diagcomM}. In addition, this map is onto the components of \eqref{diagcomM} that project to zero in \eqref{NtoM}. This means that we can combine these two diagrams into the commutative diagram:
\begin{equation}
\label{NtoMtoN}
\begin{split}
\begindc{\commdiag}[5]
	\obj(20,10)[blz]{$0$}
	\obj(30,10)[bnodes]{$\displaystyle \pi^*\mcO(i)$}
	\obj(50,10)[barr]{$\displaystyle \pi^*\mcO(a)$}
	\obj(75,10)[brel]{$\displaystyle \pi^*\mcO(-A)$}
	\obj(95,10)[brz]{$0$}
	\obj(105,10)[brrz]{$0$}
	\obj(20,20)[mlz]{$0$}
	\obj(30,20)[mnodes]{$\displaystyle  \pi^*\mcO(i)$}
	\obj(50,20)[marr]{$\displaystyle  \pi^*\mcO(a) \oplus \pi^*\mcO(A)$}
	\obj(75,20)[mrel]{$\displaystyle \pi^*\mcO(-A) \oplus \pi^*\mcO(-a)$}
	\obj(95,20)[mnodes2]{$\displaystyle \pi^*\mcO(i)$}
	\obj(105,20)[mrz]{$0$}
	\obj(20,30)[tllz]{$0$}
	\obj(30,30)[tlz]{$0$}
	\obj(50,30)[trel]{$\pi^*\mcO(A)$}
	\obj(75,30)[tarr]{$\pi^*\mcO(-a)$}
	\obj(95,30)[tnodes]{$\pi^*\mcO(i)$}
	\obj(105,30)[trz]{$0$}
	\mor{blz}{bnodes}{}
	\mor{bnodes}{barr}{}
	\mor{barr}{brel}{}
	\mor{brel}{brz}{}
	\mor{brz}{brrz}{}
	\mor{mlz}{mnodes}{}
	\mor{mnodes}{marr}{}
	\mor{marr}{mrel}{}
	\mor{mrel}{mnodes2}{}
	\mor{mnodes2}{mrz}{}
	\mor{tllz}{tlz}{}
	\mor{tlz}{trel}{}
	\mor{trel}{tarr}{}
	\mor{tarr}{tnodes}{}
	\mor{tnodes}{trz}{}
	\mor{mnodes2}{brz}{}
	\mor{mrel}{brel}{}
	\mor{marr}{barr}{}
	\mor{mnodes}{bnodes}{}
	\mor{tlz}{mnodes}{}
	\mor{trel}{marr}{}
	\mor{tarr}{mrel}{}
	\mor{tnodes}{mnodes2}{}
\enddc
\end{split}\ .
\end{equation}
This is an exact sequence in the category of complexes of sheaves. By the HKR isomorphism and flatness of $\pi$, the the middle line is isomorphic to $\bigoplus \wdg^i \pi^*TM[-i]$, the bottom line to $\bigoplus \wdg^i \pi^*TN[-i]$ and the top line to $\bigoplus \mathscr{H}om_M(\wdg^i \pi^*TN[-i],\mcO_M)[-3] \cong \bigoplus \pi^*\Omega_N^i[i-3]$. As chain maps induce maps on cohomology, and each horizontal line is quasiisomorphic to its cohomology, to show that this exact sequence splits, it suffices to show that the groups $\Ext^1_M(\pi^*\omega_N,\pi^*TM)$ and $\Ext^1(\pi^*\Omega_N,\pi^*\omega_N^{-1})$ vanish. These are both isomorphic to $H^1_M\left(\pi^*\!\left(TN \otimes \omega_N^{-1}\right)\right) \cong \bigoplus_{p\ge 1} H^1_N(TN \otimes \omega_N^{-p})$. We have assumed that this group vanishes.

It follows from the fact that we have a simple collection of exceptional sheaves \eqref{simphel} that all these sheaves are acyclic, so the functor $\Gamma(-)$ is exact. Thus, the above diagram gives rise to a commutative diagram of spaces of sections. We have from \eqref{diagcomM} and section \ref{ssec:sdres} that $\HH^*(M)$ is the cohomology of the middle line. We are interested in $\HH^2(M)$ which arises from the term $H^0(\pi^*\mcO(-A) \oplus \pi^*(\mcO(-a)))$. From the previous section, we have an embedding of $H^0(\pi^*\omega_N^{-1})$ into the cohomology of the bottom row. This is an equivalence class of elements in $H^0(\pi^*\mcO(-A))$. Pick an element of this equivalence class, 
$\sum p_A$. For each $p_A$, choose a representative $\sigma(p_A) \in \BC Q$. As $s(p_A) = t(A)$ and $t(p_A) = s(A)$, we can form the loop $\ell = \sum_A A\sigma(p_A) \in L$. From section \ref{ssec:sdres}, this defines an element in $\HH^2(M)$. It remains to show that this element is well-defined. There are two possible ambiguities. The first, the choice of an element of the equivalence class, is exactly accounted for by the image of $H^0(\pi^*\mcO(a))$ in the middle line. The second, the choice of a representative of $p_A$, is ambiguous up to an element of $I$. This is accounted for by the image of the $H^0(\pi^*\mcO(A))$.

Let us summarize what we have shown. We began with a quiver $Q$ with relations given by $I$ that was completed to a quiver $\barQ$ with relations given by the ideal $\bar{I}$ which is generated by derivatives of a superpotential $W \in L$. We have seen that an element of $H^0(\pi^*\omega_N^{-1}) \subset \HH^2(M)$ gives rise to a superpotential deformation as follows. $Q$ has an initial node, $n$. There is an embedding of $H^0(\pi^*\omega_N^{-1}) \hookrightarrow \Hom(P_n,P_n)$, the space of based loops at $n$. By choosing a representative in $\BC \barQ$, we obtain an element $\ell \in L$. Then, the deformation of the algebra given by this element in $\HH^2(M)$ is precisely given by the superpotential deformation $W \leadsto W - \hbar \ell$.

One is presented with a small puzzle now. While there is a distinguished initial node in $Q$, the nodes of $\barQ$ only have a cyclic ordering. Given that the physical information is contained in $\barQ$, what is the significance of our choice of a node? The answer is that the specific choice of an equivalence of categories between $\mcD(M)$ and $\mcD(\barQ / \bar{I})$ depends on the exceptional collection which determines for us both the quiver $Q$ and the initial node, $n$. The other quivers that give rise to the quiver $\barQ$ correspond to mutating the rightmost element of the exceptional collection to the far left. This exceptional collection gives rise to a different set of fractional branes which are related to the original fractional branes by an autoequivalence of the derived category. This theorem and others about the relation between mutation of the exceptional collection and the equivalence of categories can be found in \cite{BridgeTStruct} .

\section*{Acknowledgments}

I would like to thank David Ben-Zvi, Jacques Distler, Dan Freed and Uday Varadarajan for many useful conversations. I would also like to thank Andrei C{\u{a}}ld{\u{a}}raru, Harm Derksen and Chris Herzog for e-mail conversations. This material is based upon work supported by the National Science Foundation under Grant No. PHY-0455649. Most of this work was completed while I was a postdoctoral fellow at the University of Texas at Austin. The final stages of this work were supported by Texas A\&M University.

\bibliographystyle{utphys}
\bibliography{thebib}
\end{document}